\documentclass[journal]{IEEEtran}

\usepackage{cite}
\usepackage{epsfig,subfigure}
\usepackage{float}
\usepackage{algorithm}
\usepackage{algpseudocode}
\usepackage{amsmath}
\usepackage{color}
\usepackage{epstopdf}
\usepackage{multirow}
\usepackage{multicol}
\usepackage{arydshln}
\usepackage{graphicx}
\usepackage [latin1]{inputenc}
\begin{document}

\title{ Vehicular Edge Computing and Networking: A Survey}

\author{Lei Liu,~\IEEEmembership{Student Member,~IEEE,}
        Chen Chen,~\IEEEmembership{Senior Member,~IEEE,}
        Qingqi Pei,~\IEEEmembership{Senior Member,~IEEE,}
        Sabita Maharjan,~\IEEEmembership{Member,~IEEE,}
        and~Yan Zhang,~\IEEEmembership{Senior Member,~IEEE}
\thanks{ Lei Liu, Chen Chen and Qingqi Pei are with the State Key Laboratory of Integrated Service Networks, Xidian University, Xi'an 710071, China(e-mail:tianjiaoliulei@163.com;cc2000@mail.xidian.edu.cn;
qqpei@mail.xidian.edu.cn).}
\thanks{ Sabita Maharjan is with Simula Metropolitan Center for Digital Engineering
and University of Oslo, Norway (email: sabita@simula.no).}
\thanks{ Yan Zhang  is with the Department of Informatics,
University of Oslo, Norway (email: yanzhang@ieee.org).}
\thanks{Manuscript received XX, 2018; revised XX, 2018.}}

\markboth{Journal of \LaTeX\ Class Files,~Vol.~XX, No.~XX, XX~2018}
{Shell \MakeLowercase{\textit{et al.}}: Bare Demo of IEEEtran.cls for IEEE Journals}

\maketitle

\begin{abstract}
As one key enabler of Intelligent Transportation System (ITS), Vehicular Ad Hoc Network (VANET) has received remarkable interest from academia and industry. The emerging vehicular applications and the exponential growing data have naturally led to the increased needs of communication, computation and storage resources, and also to strict performance requirements on response time and network bandwidth. In order to deal with these challenges, Mobile Edge Computing (MEC) is regarded as a promising solution. MEC pushes powerful computational and storage capacities from the remote cloud to the edge of networks in close proximity of vehicular users, which enables low latency and reduced bandwidth consumption. Driven by the benefits of MEC, many efforts have been devoted to integrating vehicular networks into MEC, thereby forming a novel paradigm named as Vehicular Edge Computing (VEC). In this paper, we provide a comprehensive survey of state-of-art research on VEC. First of all, we provide an overview of VEC, including the introduction, architecture, key enablers, advantages, challenges as well as several attractive application scenarios. Then, we describe several typical research topics where VEC is applied. After that, we present a careful literature review on existing research work in VEC by classification. 
Finally, we identify open research issues and discuss future research directions.

\end{abstract}


\IEEEpeerreviewmaketitle

\section{Introduction}

\begin{figure*}
  \centering

  \includegraphics[height=6cm, width=11cm]{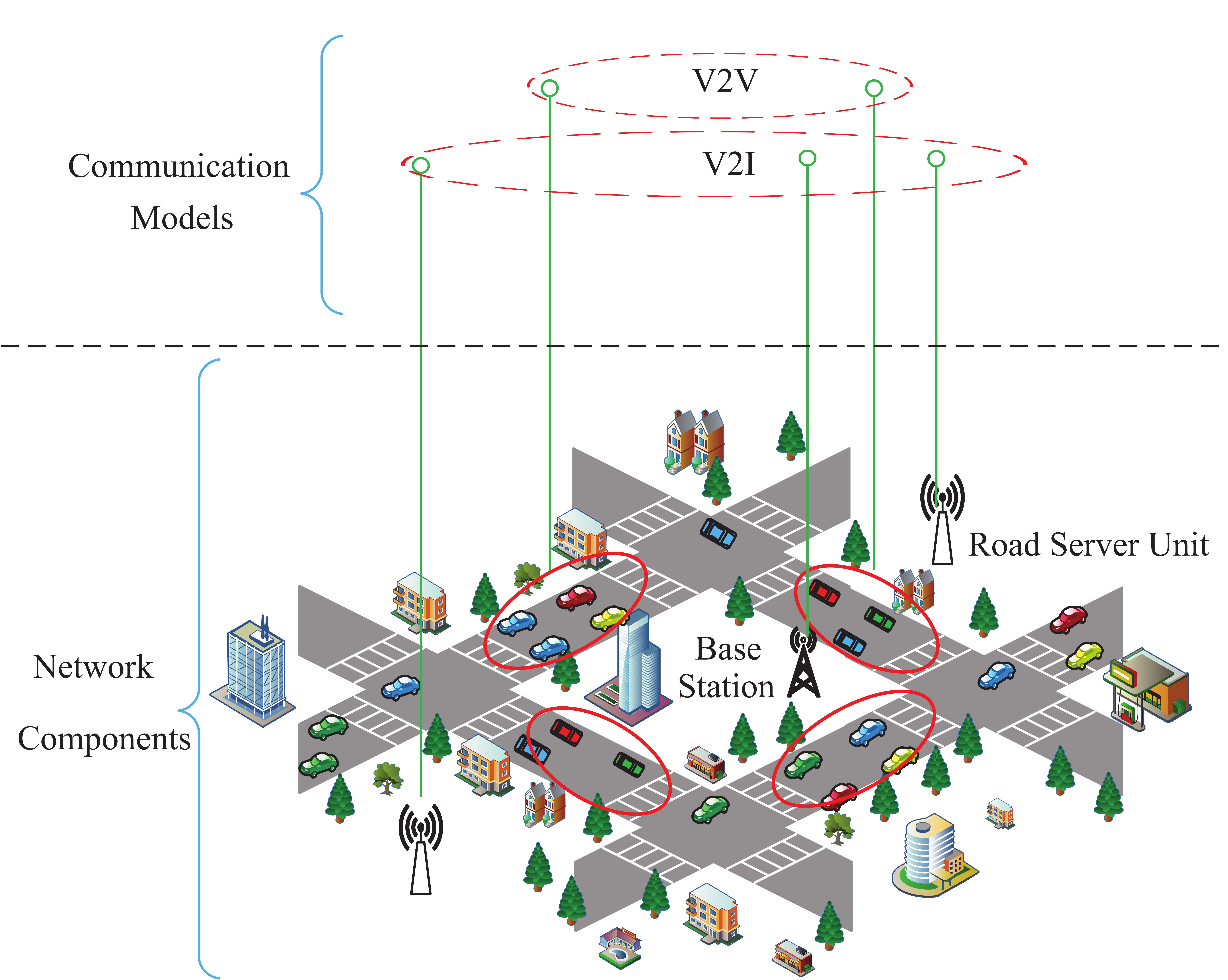}
  \caption{Architecture of vehicular networks}
  \label{f1}
\end{figure*}

 Due to the rapid pace of industrialization and urbanization, Intelligent Transportation System (ITS) has received increasing interest from academia and industry. By integrating information and communication technologies \cite{lien2019latency}, \cite{li2017secure}, \cite{li2018uav},\cite{mao2019power}, \cite{zhai2018energy}, \cite{zhang2011home}, ITS plays an important role in enhancing road safety and improving traffic efficiency. Vehicular Ad Hoc Network (VANET)\cite{peng2018vehicular} is a key enabler in ITS. Being a special kind of Mobile Ad Hoc Networks (MANETs), VANETs are comprised of two basic elements:vehicles and Road-Side Units (RSUs)\cite{lien2018low}, \cite{chen2018asgr}, as shown in Fig. \ref{f1}. Vehicles are equipped with communication devices, which enables short-range wireless transportation.  RSUs are distributed along the road to be connected to the backbone network for the purpose of facilitating network access. Data communication in VANETs can be realized in two models: Vehicle-to-Vehicle (V2V) and Vehicle-to-RSU (V2R) \cite{kaiwartya2016internet}, \cite{jinna2018cvcg}. Using the two communication models, vehicular networks support an array of applications, which include three main categories: 1) road safety applications (e.g., lowering the risk of accidents); 2) traffic efficiency applications (e.g., reducing travel time and alleviating traffic congestion; 3) value-added applications ( e.g., providing infotainment, path planning and internet access).

The rapid development of vehicular networks will facilitate wide use of smart vehicles, which enables a large number of various types of applications \cite{zhou2018dependable},\cite{zhou2018social}. The implementation of these applications call for enormous resources for data storage and processing. Constrained by the limited computation and communication capacities, vehicles cannot readily satisfy the increasing resource demands of such applications, specially those with intensive computation and strict delay requirements. 
In order to tackle these issues, Mobile Cloud Computing (MCC) is widely regarded as a promising solution \cite{dinh2013survey}. With the combination of computation and communication technologies, MCC allows to run application services of users at the remote cloud. Thus, MCC can bring the following benefits for users: 1) reduce energy consumption; 2) enable sophisticated services; 3) provide huge storage capacity. There have been several survey articles on MCC with different focuses \cite{dinh2013survey}, \cite{rahimi2014mobile}, \cite{alizadeh2016authentication}. In \cite{dinh2013survey}, the definition, architecture and application of MCC are introduced, and the existing issues and corresponding approaches are presented. The work in \cite{rahimi2014mobile} discusses  applications, challenges and opportunities of MCC. The taxonomy and comparison of state-of-the-art authentication schemes regrading MCC are presented in \cite{alizadeh2016authentication}. In order to enhance road safety and improve travel comfort, significant amount of effort has been denoted to integrating MCC with vehicular networks \cite{lin2018resource}. The authors in \cite{lin2013cloud} propose a cloud-supported gateway model to seamlessly access the internet in ITS. By this model, the user experience can be improved. A VANET-cloud computing model is devised in \cite{bitam2015vanet}, where the cloud computing resources are used to enhance the quality of service (QoS). Depending on the presented cloud system in \cite{mershad2013finding}, vehicles can find their requested resources from mobile mobile services.

Despite the advantages of MCC, considering that the cloud is far from users, high transmission latency is a consequence. Moreover, the explosive growth of mobile data will further impose huge stress on the load of back-haul networks. 
If all the data are sent to the cloud for processing,  the bandwidth consumption and competition are extremely significant. Mobile Edge Computing (MEC) interchangeably regarded as fog computing (FC), is envisioned as a promising paradigm to address such issues \cite{wang2017survey},\cite{wang2016mobile}, \cite{dai2018joint}, \cite{du2019enabling}. In MEC, the cloud service is pushed to the network edge, i.e., the computation and storage resources are moved to  proximity of users, by which latency can be greatly reduced and energy can be saved to a large extent. 
MEC literature has also seen several survey papers \cite{abbas2018mobile},\cite{mao2017survey}, \cite{roman2018mobile},\cite{taleb2017multi}. The work in \cite{abbas2018mobile} gives a comprehensive overview of existing research developments in MEC along with  advantages, architectures and applications. Meanwhile, the issues and existing solutions of security and privacy are also elaborated.  The authors in \cite{mao2017survey} introduce the work on computing and communication in MEC with a focus on joint-radio-and-computational resource allocation. 
The enabling technologies in MEC, such as Virtual Machine (VM), Software Defined Networking (SDN), Network Function Virtualization (NFV) are discussed in \cite{taleb2017multi}.

Vehicular Edge Computing (VEC) has a great potential to enhance traffic safety and improve travel comport by integrating  MEC into vehicular networks. VEC literature has seen several work mainly on computation offloading \cite{qiao2018collaborative}, \cite{du2015computation}, content caching \cite{hou2018q},\cite{guo2018cache}, security and privacy \cite{cui2018efficient} and\cite{huang2017distributed}. In view of prior work, it can be seen that 
there has been no work to survey the existing research in VEC. In this paper, we make an attempt to  present a comprehensive  review of state-of-the-art research  on VEC
. The contributions of this paper are summarized as follows:

\begin{itemize}
  \item A comprehensive overview of VEC is provided, where the introduction, architecture, advantages, challenges and application scenarios of VEC are elaborated in detail.


  \item The existing research models in VEC, which are related to computation offloading, content caching, data management, flexible network management as well as security and privacy are summarized, respectively. Besides, a comprehensive literature review for each model is presented.

  \item The open issues are identified and future research directions are discussed.

\end{itemize}

\begin{figure*}
  \centering

  \includegraphics[height=3cm, width=16cm]{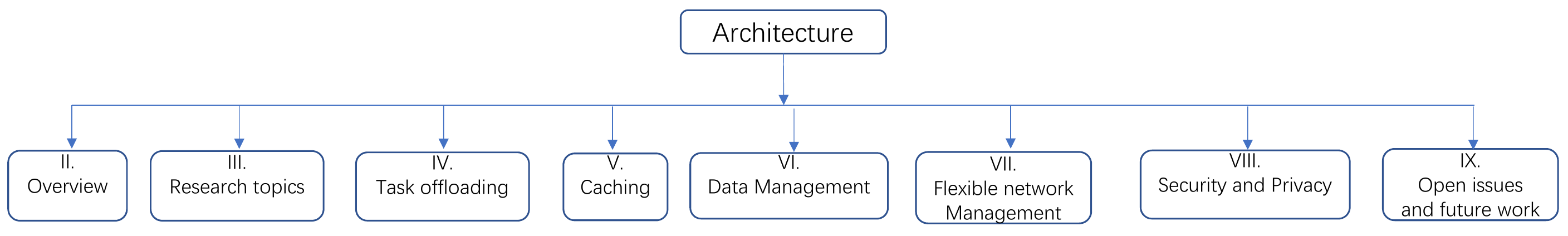}
  \caption{Architecture of the content organization}
  \label{f01}
\end{figure*}

The organization of this paper is illustrated in Fig. \ref{f01}, which is introduced as follows. We provide an overview of VEC in Section \ref{s2}. We present  several VEC research topics in Section \ref{s3}. In Section \ref{s4}-\ref{s8}, we summarize the existing elaborated models in VEC, respectively. The open issues and future research directions are introduced in Section \ref{s10}, followed by the conclusion in Section \ref{s11}. The used definitions of acronyms in this paper is summarized in Table I for easy reading.

\begin{table}

\caption{SUMMARY OF ACRONYMS}
\begin{center}
\begin{tabular}{|l|l|}
\hline
\textbf{Acronym} & \textbf{Definition}  \\ \hline
ITS &Intelligent Transportation System\\
\hline
VANET&Vehicular Ad Hoc Network\\
\hline
MANET&Mobile Ad Hoc Network\\
\hline
RSU&Road-Side Units\\
\hline
V2V&Vehicle-to-Vehicle\\
\hline
V2R&Vehicle-to-RSU\\
\hline
MCC&Mobile Cloud Computing\\
\hline
QoS&Quality of Service\\
\hline
MEC&Mobile Edge Computing\\
\hline
FC&Fog Computing\\
\hline
VM&Virtual Machine\\
\hline
SDN&Software Defined Networking\\
\hline
NFV&Network Function Virtualization\\
\hline
VEC&Vehicular Edge Computing\\
\hline
AR&Augmented Reality\\
\hline
AI&Artificial intelligence\\
\hline
GPS&Global Positioning System\\
\hline
HD&High Definition\\
\hline
V2I&Vehicle-to-Infrastructure\\
\hline
AV&Autonomous Vehicle \\
\hline
SBS&Small Base Station\\
\hline
E2E&End-to-End\\
\hline
PVs&Parked Vehicles\\
\hline
QoE&Quality of Experience\\
\hline
UAVs&Unmanned Aerial Vehicles\\
\hline
SRSU&Solar-powered RSUs\\
\hline
P2P&Peer-to-Peer\\
\hline
BSs&Base Stations\\
\hline
LUR& Load Utilization Ratio\\
\hline
QCR &Query to Connectivity Ratio\\
\hline
MLP&Multi-Layer Perception \\
\hline
CNN&Conventional Neural Network \\
\hline
VDTNs&Vehicular Delay-tolerant Networks\\
\hline
5G-SDVNs&5G-enabled Software-Defined Vehicular Networks\\
\hline
RF&Radio Frequency \\
\hline
LOS&Line of Sight \\
\hline
VFS&Vehicular Fog Service \\
\hline
CL-A-SC&CertificateLess Aggregate SignCryption\\
\hline
POI&Point of Interest\\
\hline
\end{tabular}
\end{center}
\end{table}

\begin{figure*}
  \centering

  \includegraphics[height=8cm, width=16cm]{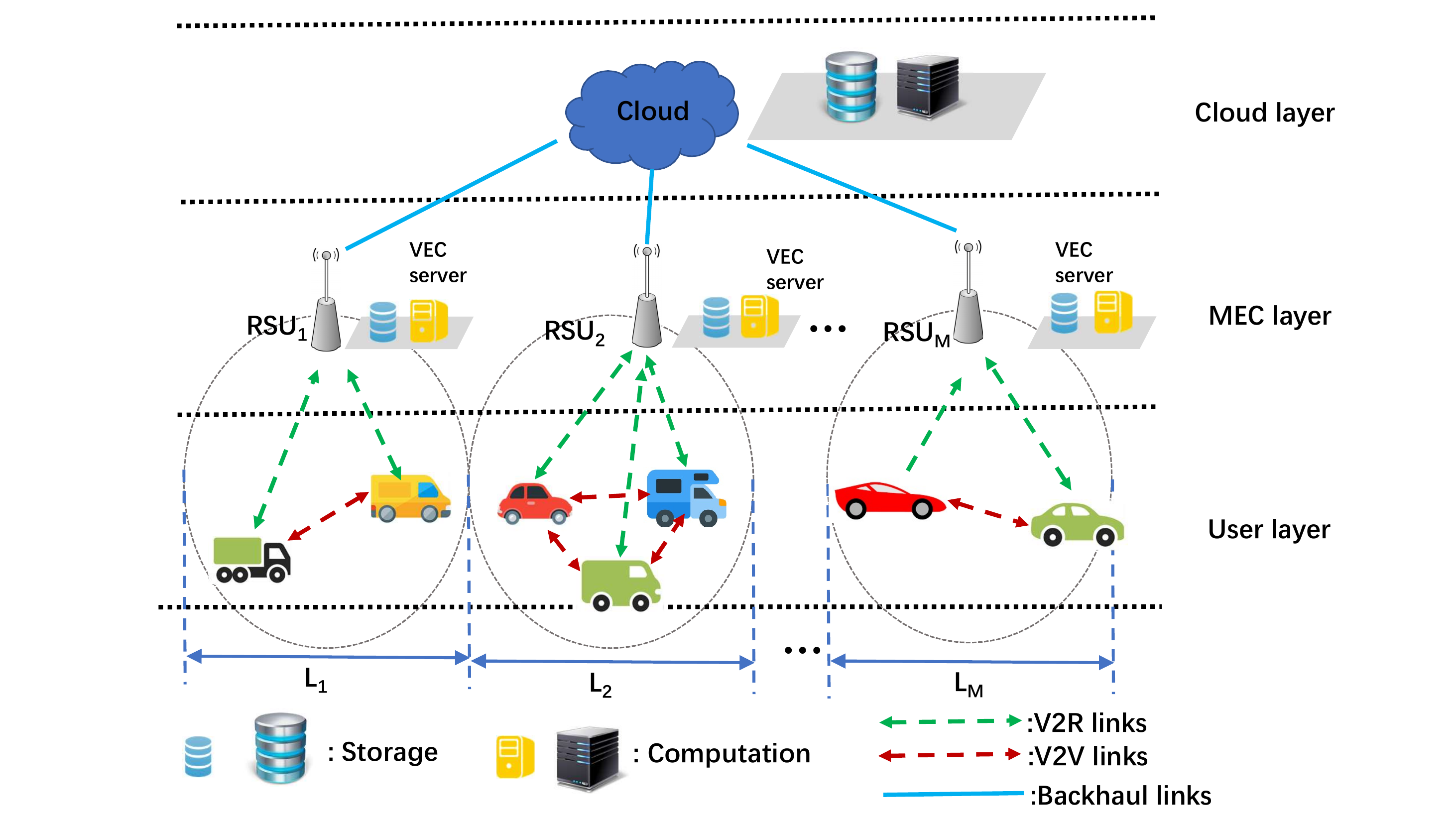}
  \caption{Architecture of vehicular edge computing}
  \label{f2}
\end{figure*}

\section{Vehicular Edge Computing: An Overview} \label{s2}

VEC is the integration of emerging MEC with traditional vehicular networks.  VEC aims to move communication, computing and caching resources to close proximity of vehicular users. Thus, VEC has the potential of playing a key role in  addressing the exponentially increasing needs of edge devices on low delay and high bandwidth. Different from traditional MEC, the prominent features of VEC are  fast mobility of vehicles which leads to the frequent and more dynamic topology changes and the complicated communication characteristics due to the rapidly varying channel environment over time. 

In VEC, vehicles own certain communication, computation and storage resources; RSUs which often act as edge servers are placed close to vehicles for gathering, processing and storing data timely. Due to the constrained capacity, vehicles can offload their computation-intensive and latency-sensitive tasks to the edge servers, which can considerably reduce the response time and efficiently alleviate the heavy burden on backhaul networks. 
In a VEC framework, the content requester can directly obtain the needed content from caching nodes without accessing the core network, thus reducing end-to-end latency and improving efficiency of network bandwidth usage. Due to the limited storage space, each caching node cannot cache all the contents. Thus, the following key issues arise: how to determine what to cache, how to determine where to cache and how to determine cache policy.

It is anticipated that by combing emerging new technologies, such as SDN \cite{nunes2014survey}, blockchain \cite{puthal2018everything}, Artificial Intelligence (AI) \cite{wang2015artificial},\cite{lv2018deep}, \cite{yue}, \cite{yu2018svms},\cite{guo2019deep}, VEC enable services  efficiently and effectively.

Next, we introduce architectures, advantages, challenges and application scenarios of VEC, respectively.

\subsection{Vehicular Edge Computing:Architecture}

The architecture of VEC is  presented in Fig. \ref{f2}. Typically, this architecture is composed of three layers, i.e., vehicular terminals as the user layer, RSUs as the MEC layer and cloud servers as the cloud layer.
\subsubsection{Vehicular Terminals}

In VEC, vehicular terminals are mainly represented by vehicles. Rather than ordinary mobile nodes, vehicles have the following prominent features: 1) sensing: vehicles can sense the environment from both inside and outside and are able to collect various information using the equipped vehicular devices, including cameras, radars, Global Positioning System (GPS), etc.; 2) communicate: vehicles can exchange and share information with other vehicles or RSUs using V2V and V2R communication manners; 3) computing: in addition to transferring parts of computation tasks to the edge servers or the cloud for processing, vehicles can execute parts of the tasks locally by themselves; 4) storage: the idle storage space of vehicles can be used to cache popular contents for data sharing.

\subsubsection{Edge Servers}

RSUs often act as edge servers in VEC, which are  distributed along the road in a city.  They have rich communication, computation and storage resources compared to vehicles. RSUs are responsible for receiving the information sent from vehicles, processing these collected information, and even uploading these information to the cloud. By computation offloading and caching technologies, RSUs are beneficial for handling strict performance requirements. Besides, they can also provide diverse services for vehicles, such as  video streaming, traffic control, path navigation.

\subsubsection{Cloud Servers}

Cloud services are deployed in a remote cloud.  They can get the uploaded information from edge servers. Compared with edge servers, cloud services have rich capacities in terms of computation and storage, and cover a much broader area. By getting the uploaded information from mobile nodes and edge servers, they can have a global view of the covered area. The cloud paradigm can provide global level management and centralized control, which helps in making optimal decisions.

\subsection{Vehicular Edge Computing:Key Enablers }

In this subsection, several enabling technologies for VEC are introduced, including cloud technology, SDN \cite{luo2019simplifying}, NFV and smart vehicles \cite{wang2017survey}, \cite{abbas2018mobile}, \cite{mao2017survey}.

\subsubsection{Cloud Technology  \cite{rahimi2014mobile}}

Cloud has powerful computation capacity and storage resources. Generally speaking, it is deployed far away from the users, which will incur the long latency. By moving the functionality of the cloud to the network edge,  VEC is capable of sustaining diverse application services.

\subsubsection{Software Defined Networking \cite{nunes2014survey}}

SDN is regarded as an innovation network architecture, the root feature of which is to decouple the control plane and the data plane. By this method, it can simply the network management.

\subsubsection{Network Function Virtualization  \cite{mijumbi2016network}}

NFV aims to provide a new approach to design, deploy as well as manage network services. Its key is to decouple network functions from the physical devices where they run. It can significantly reduce the operating costs and capital costs as well make the deployment of new services more flexible and efficient.

\subsubsection{Smart Vehicles}

With the development of vehicular equipments and information technologies, vehicles will become much smart in the future. At that moment, smart vehicles not only communicate with each other, but also have own ample resources in terms of computation and storage. Using these resources, vehicles can process tasks locally, alleviating the network burden and reducing the delay.

\subsection{Vehicular Edge Computing: Advantages}

The key advantages of VEC are as follows.

\begin{figure*}
  \centering

  \includegraphics[height=8cm, width=16cm]{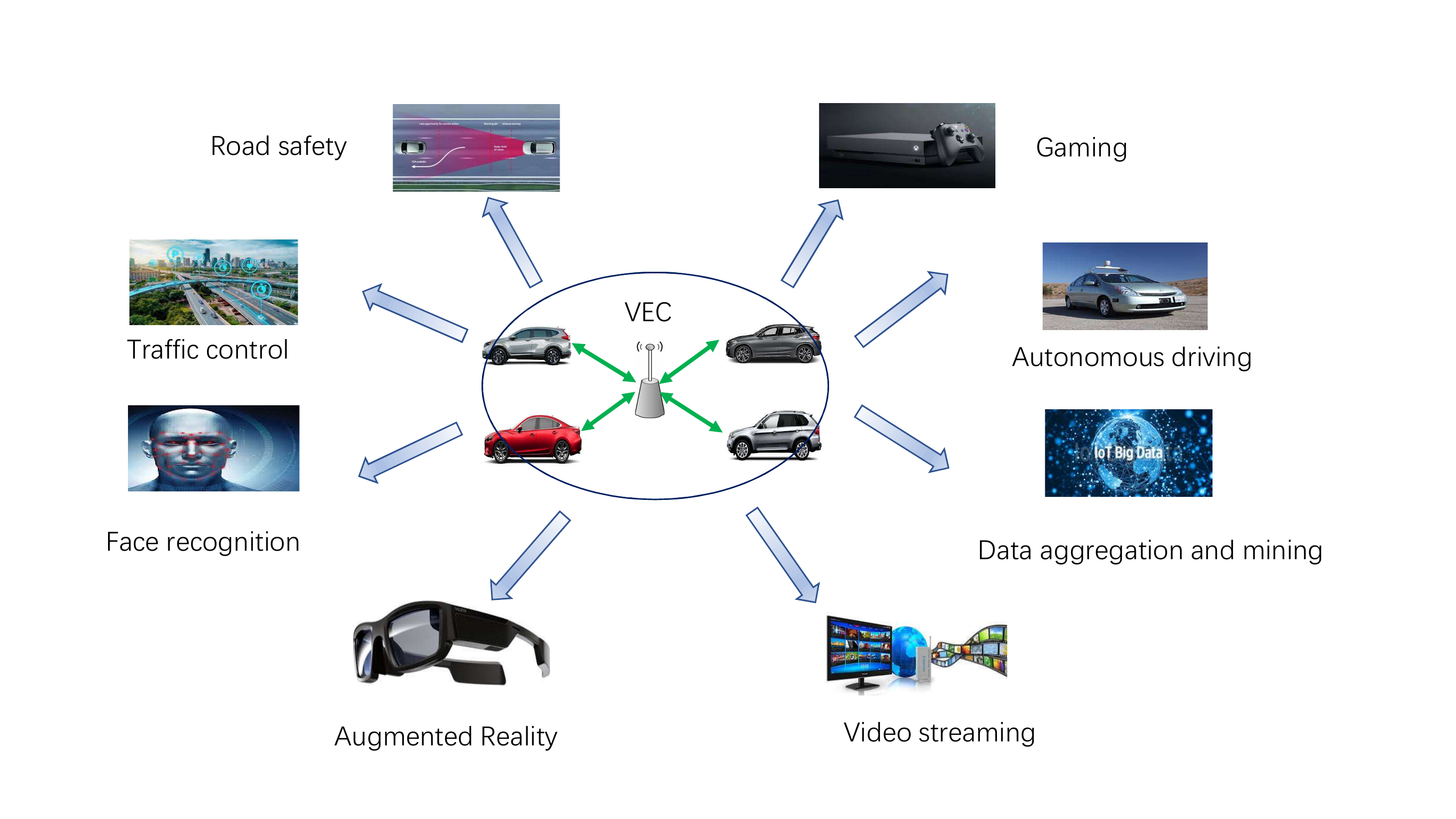}
  \caption{Application scenarios of vehicular edge computing}
  \label{f3}
\end{figure*}

\subsubsection{Response Time}

The response time consists of the delivery time of data offloaded to edge servers and back plus the time for processing in the servers. When compared with the cloud, the edge servers in VEC  are closer to vehicular users. Thus, the took execution time is considerably less, which is specially beneficial for delay-sensitive applications, such as safety applications.

\subsubsection{Energy Efficiency}

The increasing prevalence of smart vehicles drives an explosive growth of various vehicular applications, ranging from driving, communication, and computing to storage in the future. These applications will bring huge energy consumption. Assisted by VEC, electric vehicles with limited energy can still provide enough support for these applications.

\subsubsection{Bandwidth}

With the popularization of smart vehicles, the amount of generated data from them will grow explosively. In addition, the content requests will also become diverse. Because of the long distance from users, cloud computing cannot guarantee the bandwidth requirements for processing such enormous amount of data computation and content delivery using the centralized management. By moving the computation and storage resources of the cloud to the network edge, VEC can efficiently alleviate the huge bandwidth stress to back-haul networks.

\subsubsection{Storage}

Different from the cloud, the data can be stored in edge servers which are in proximity of vehicular users in VEC. The caching technology enables to access the stored data in time for vehicular users, and reduces the storage burden on the remote cloud.

\subsubsection{Proximity Services}

Because edge servers are closer to vehicular users in VEC, various proximity services can be provided to the users. In this way, the user experience can be better guaranteed while maintaining efficiently  the traffic management. For example, when receiving the location and sensing information uploaded from vehicles, edge servers can help in processing the data to build a high definition (HD) map and then send the HD-map to vehicles.

\subsubsection{Context Information}

 Edge servers in VEC can obtain the real time information related to the behavior and location of vehicle, traffic environments, network conditions, etc. With these information, various applications can be improved. One example is that these real time information can be used to deliver contents to vehicular users based on their interests.

\subsection{Vehicular Edge Computing:Challenges}

The main VEC challenges are as follows.

\subsubsection{High Mobility}

Due to the high mobility of vehicles,  the network topology in vehicular environments is highly dynamic changing \cite{al2014comprehensive}. Under this condition, links can be easily disconnected, which will deteriorate  communication quality. Besides, vehicles  may switch among multiple edge servers, leading to handovers. 
The frequent handovers contribute to the delay and inversely impact service continuity, degrading the user experience.  

\subsubsection{Harsh Channel Environment}

In vehicular networks, especially urban vehicular environments, there exist many obstacles, e.g., trees, buildings, which may hinder the success of data transmission \cite{sharef2014vehicular}. Meanwhile, it is difficult to characterise the time-varying channels.

\subsubsection{Resource Management}

Compared to cloud computing, the resources of VEC in terms to computation and storage are limited. Thus, how to manage these resources is vital. Given dynamic resource demands, diverse application characteristics, complicated traffic environments, the optimization of resource allocation is a challenging task.

\subsubsection{Task Migration}

Due to the constrained capacity, it is needed for vehicular users to offload computation-intensive and delay-sensitive tasks to  edge servers. Considering the dynamic channel environment and frequent changing topology, optimizing task migration decision is crucial.
\subsubsection{Security and Privacy}

In VEC, due to the dynamic topology changes, vehicles may not fully trust each other. 
Meanwhile, when the same physical edge servers are allowed to be accessed by different vehicular users, there are the issues of security and privacy without a strong protection mechanism.

\subsection{Vehicular Edge Computing:Application Scenarios}
The emergence of VEC enables diverse types of applications as illustrated in Fig. \ref{f3}. In the following, several typical scenarios where VEC can be applicable are introduced.

\subsubsection{Road Safety}
Receiving directly the data forwarded from vehicles and sensors deployed along the road, edge servers in close proximity of vehicles can real-time analyze and process these data. Once finding risk data, edge servers are responsible for notifying surrounding vehicles to avoid possible danger by taking reasonable measures, such as braking, lane changing and turning round.

\subsubsection{Entertainment}
With the advent of smart vehicles, drivers can be liberated from complex driving operations and spend abundant time on entertianment, e.g., surfing the internet, playing gaming, watching video \cite{chen2019study}. These applications will benefit from rich computation and storage resources in VEC. For example, by caching popular contents cooperatively among edge servers and vehicles, vehicular users can directly fetch the requested videos without resorting the remote cloud, thereby decreasing the delay and improving the experience of users.

\subsubsection{Traffic Control}

In VEC, edge servers cover a communication region. Each vehicle within this region send its current status (location, speed) and collected information (weather conditions, road conditions) to the associated server. After receiving the information uploaded from vehicles, the server has a knowledge of local traffic conditions, then it can control traffic flow. By this method, traffic congestion \cite{chen2017congestion} can be avoided.

\subsubsection{Path Navigation}

Real-time navigation system is of quite importance to provide an optimal route for drivers. The implementation of real-time navigation involves data sensing, collection and processing. VEC is capable of providing significant computation and storage resources for real-time navigation system.

%

\subsubsection{Ultra-low Latency Service}

VEC allows service provisioning with ultra-low latency and high reliability, e.g., autonomous driving \cite{yuan2018toward}. 
In autonomous driving, it is crucial for vehicles to understand surrounding environments accurately and timely and performs more efficient and effective operations. VEC makes it possible for autonomous driving to obtain the support of powerful computation resource for task execution. 



\subsubsection{Computation-intensive Service}

The computation offloading of VEC can be leveraged to service computation-intensive applications. Two typical examples among these applications are Augmented Reality (AR) and face recognition. Due to the limited resource, vehicular users cannot satisfy their computation requirements. These applications can be migrated to edge servers empowered by powerful computation resource in VEC for implementation.

\subsubsection{Data Aggregation and Data Mining}

The prevalence of smart vehicles drives the exponential growth of data. These data can be generated by sensors or received from other vehicles \cite{zhang2018artificial}. If these data are be deeply exploited in VEC, more learned knowledge can be used to facilitate the data efficiency as well as improve the network performance.

\section{Research Topics in Vehicular Edge Computing}\label{s3}

There have been many efforts devoted on the research of VEC,  mainly including the following topics :

\begin{enumerate}
  \item Task offloading: In addition to be processed locally, tasks can be offloading to other edge devices for processing.
  \item Caching: The surplus storage resources of edge devices can be exploited for content caching.
  \item Data management: Data management involves data collecting, processing and dissemination, etc.
  \item Flexible network management: Several new technologies, e.g., SDN, can be integrated into VEC to simply network management.
  \item Security and privacy: The characteristics of vehicular networks bring new challenges on security and privacy in VEC compared to MEC.
\end{enumerate}

In the following, we will introduce the research topics stated above in detail from Section \ref{s4} to Section \ref{s8}, respectively.


\section{Task Offloading in Vehicular Edge Computing}\label{s4}

\begin{figure}[H]
  \centering
  \includegraphics[height=4cm, width=8cm]{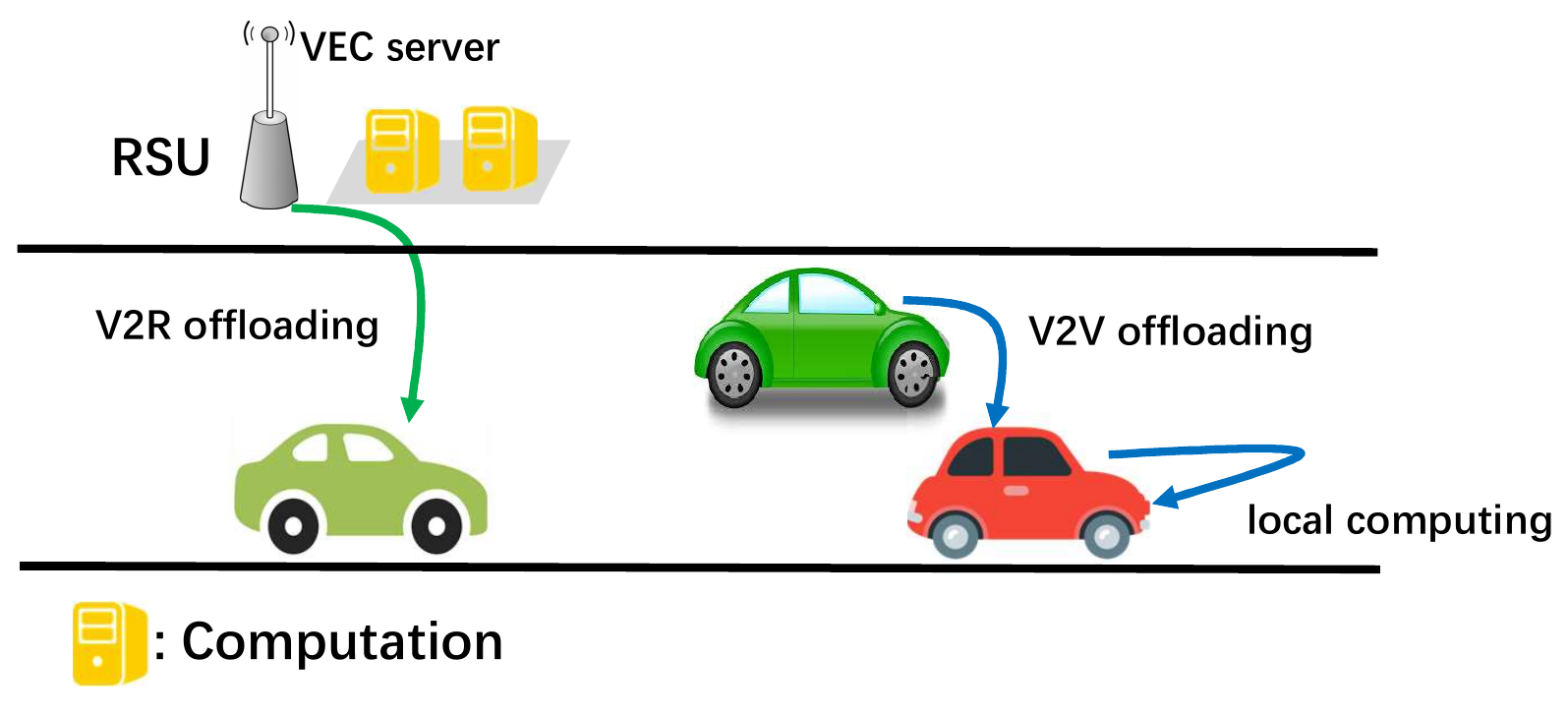}
  \caption{Example of task offloading }
  \label{f41}
\end{figure}

With the rapid deployment of vehicular networks, more and more vehicles become smart. The proliferation of smart vehicles will arise a significant increase of various types of application services. Many of these application services  have stringent requirement in terms of computation and delay. Despite abundant resources, the cloud is unfeasible to support delay-sensitive application services because of the long distance from vehicular users. Besides, massive data processing generated from these application services will bring huge pressure on the network bandwidth.  By contrast, VEC is envisioned as a promising solution. As shown in Fig. \ref{f41},  in addition to being  executed locally in VEC, user tasks can be transferred to the network edge empowered by the cloud with respect to computation and storage resources, greatly shortening the delay and alleviating the network load.

\begin{figure}[H]
  \centering
  \includegraphics[height=6cm, width=8cm]{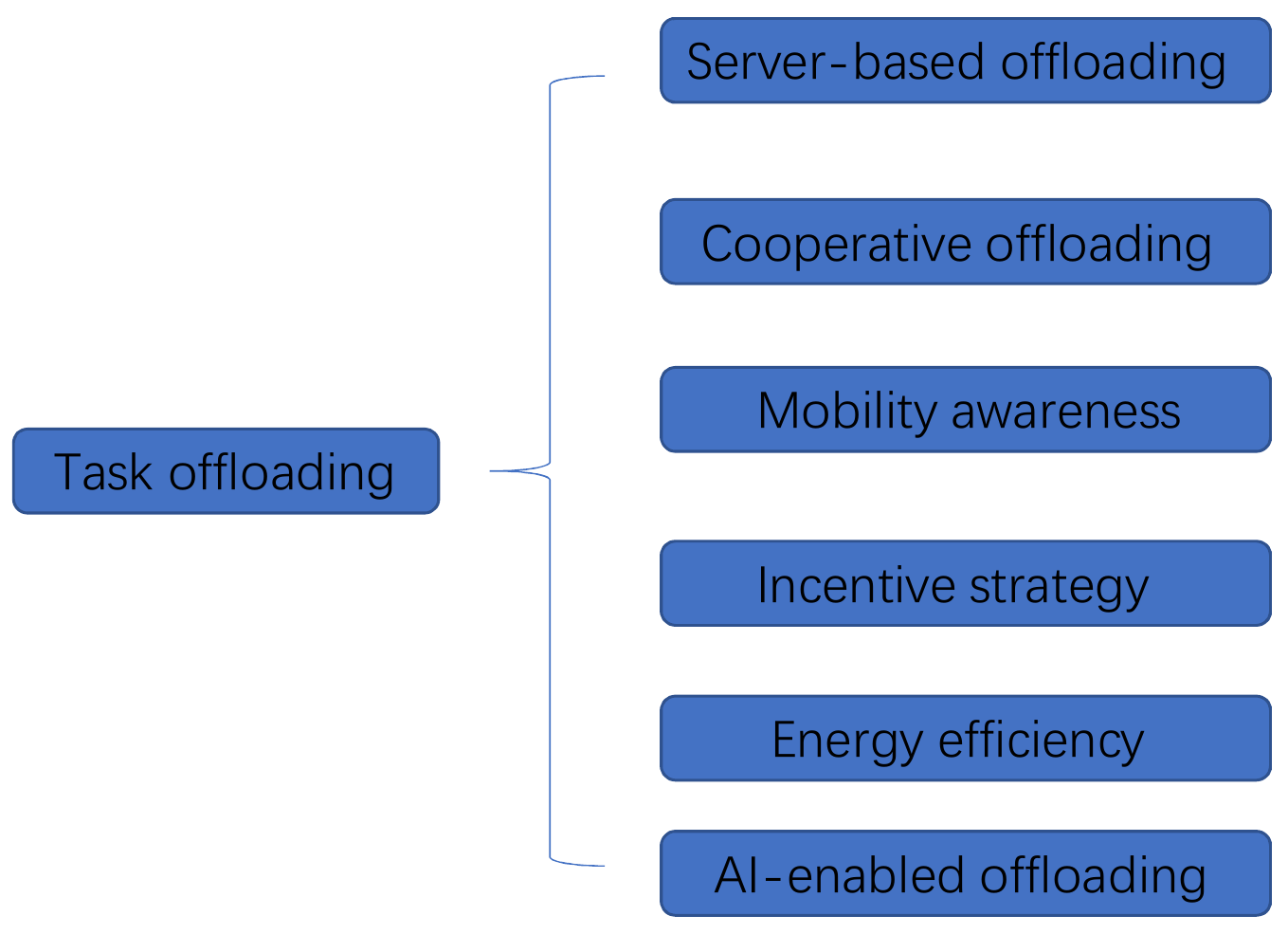}
  \caption{Classification of Offloading models}
  \label{f5}
\end{figure}

A lot of research has been done to investigate task offloading in VEC. In this section, some existing work on task offloading in VEC are introduced by classification, as shown in Fig. \ref{f5}. A summary of literatures on task offloading is illustrated in Table II.

\begin{table*}

\caption{Summary of literatures on task offloading}
\begin{center}
\begin{tabular}{|p{3cm}|p{1cm}|p{3cm}|p{8cm}|}
\hline
\textbf{Theme} & \textbf{Reference} & \textbf{Metric}& \textbf{Key Points} \\ \hline
\multirow{4}*{Server-based Offloading} &\cite{liu2018computation}& Delay& Propose a multi-user computation offloading scheme\\
\cline{2-4}
&\cite{du2018computation}& Cost & Present a dual-side optimization problem to minimize the costs of vehicles and corresponding edge server simultaneously\\
\cline{2-4}
 &\cite{tareq2018ultra}&Reliability& Propose a scheme to jointly make the optimization of AV-to-SBS association and bandwidth allocation\\ \hline
\multirow{6}*{Cooperative Offloading}&\cite{huang2018parked}&Cost&The resources of PVs are exploited for task executation\\ \cline{2-4}
& \cite{li2018parked}&Utility&An  incentive scheme based on contact theory  is designed to encourage PVs to share their resources\\ \cline{2-4}
&\cite{qiao2018collaborative}&Delay & present a collaborative task offloading scheme\\ \cline{2-4}
&\cite{sun2018task}&Delay& A learning-based task replication scheme is designed with combinatorial multi-armed theory \\ \cline{2-4}
&\cite{lin2018novel}&Utility& The designed network system can improve the QoE of users while satisfying the QoS of users\\ \cline{2-4}
& \cite{zhu2018fog}&Utility&  Propose a task allocation optimization problem assisted by static and mobile fog nodes \\\hline
\multirow{2}*{Mobility Awareness}&\cite{zhang2017mobile}&Cost&  The tasks can be adaptively offloaded to edge servers through two methods:directly offloading and predictive transmission\\ \cline{2-4}
&\cite{dai2018joint2}&Utility&The offloading scheme accounts for load balancing \\ \hline
\multirow{2}*{Incentive Strategy}&\cite{zhang2017contract}&Utility&Contract theory is utilized to determine optimal offloading strategies\\ \cline{2-4}
&\cite{zhang2017optimal} &Utility& Stackelberg game is leveraged to devise a multi-level offloading algorithm\\ \hline
\multirow{3}*{Energy Efficiecy}&\cite{zhou2018energy}&Energy& The energy consumption optimization problem is formulated as a joint workload offloading and power control problem\\ \cline{2-4}
&\cite{zhang2018energy}&Utility&UAV is introduced to improve computation performance due to the low cost and flexible deployment\\ \cline{2-4}
&\cite{ku2018quality}&QoS&Optimize the QoS of users with the assistance of solar-powered RSUs\\ \hline
\multirow{2}*{AI-enabled Offloading}&\cite{he2018integrated}&Revenue& Networking, caching and computation are jointly considered to improve the network performance\\ \cline{2-4}
&\cite{tan2018mobility}&Revenue& Present a  resource allocation strategy with the consideration of  high mobility of vehicles and strict delay requirement\\ \hline
\end{tabular}
\end{center}
\end{table*}

\subsection{Server-based Offloading}

In VEC, edge servers (e.g., RUSs) play an important role in facilitating the network functionality of edge computing evolving from the cloud computing. Edge servers are common places selected by vehicular users for task offloading, for the reason that they can process the tasks effectively without seeking help from the remote cloud. Besides communicating with other devices and acting as gateways to access the internet, RUSs in vehicular networks can be enhanced and upgraded by rich computing and storage resources for service provisioning. The work in \cite{liu2018computation} investigates the problem of multiple-user computation offloading on an edge server with the aim to reduce communication overhead. The computation offloading of each vehicle refers to channel selection for uploading its task to the edge server. Considering the coupling of offloading decision from vehicles, game theory is used to make optimal offloading decisions and suitable channel selection for vehicles. When determining the offloading strategy, the authors of \cite{du2018computation} take into account the profits of both vehicle and edge server simultaneously. To the end, a dual-side optimization is formulated to minimize their costs. At the side of vehicles, the offloading decision and local CPU frequency are jointly optimized; while at the side of edge server, the radio resource allocation and service provisioning are considered at the same time. Different from the studies in  \cite{liu2018computation} and \cite{du2018computation} with a focus of single edge server, multi-server scenario is considered in \cite{tareq2018ultra}, where a high reliability and low latency Vehicle-to-Infrastructure (V2I) communication architecture is proposed by jointly optimizing Autonomous Vehicle (AV)-to-Small Base Station (SBS) association and wireless resource management. Using  matching theory, the provided optimization problem accounts for not only  End-to-End (E2E) delay demands of AVs but also scarce wireless bandwidth resources and limited computational resources of SBSs.

\subsection{Cooperative Offloading}

With the ever-increasing application demands, more communication, computation and storage resources are required urgently. Considering the scarcity of computation and bandwidth resources of edge servers as well as the high cost of deploying these edge servers, it is necessary to utilize all available underutilized resources from existing network entities. By aggregating abundant and unused resources of vehicles, the QoS of applications can be greatly enhanced \cite{hou2016vehicular}. Because of rich and idle resources, Parked Vehicles (PVs) have the potential of providing a powerful platform for service provisioning.  The resources of PVs are fully exploited in \cite{huang2018parked} to coordinate with edge servers for running application services. A response scheduling algorithm based on Stackelberg game is formulated to implement resource allocation among PVs with the objective of reducing the overall cost of users. The other investigation of exploring the resources of PVs has been done in \cite{li2018parked}, where an incentive scheme is designed to provide rewards for PVs when their resources are shared.

On the other hand, some vehicles with unused resources can be leveraged to assist other vehicles in computation offloading. Vehicles are treated as rich resources in \cite{qiao2018collaborative}
. A collaborative task offloading strategy is presented \cite{qiao2018collaborative} to execute tasks in a cooperative manner with the goal of achieving low computation and communication delay. The work in \cite{sun2018task} targets the exploitation of ample computation resources on vehicles as well as the improvement of QoS by the proposed learning-based task replication scheme. In this scheme, vehicles are scheduled to process the same task replication at the same time. In \cite{lin2018novel},  resources from different devices are cooperatively scheduled to improve the Quality of Experience (QoE) of users and enhance the QoS of service providers. A novel task allocation solution proposed in \cite{zhu2018fog}  takes into consideration the mobility of vehicles to strive for a balance between latency and quality, where vehicles are classified into two types: the one generating data and the one acting as fog node. 

\subsection{Mobility Awareness}

\begin{figure}[H]
  \centering
  \includegraphics[height=4cm, width=8cm]{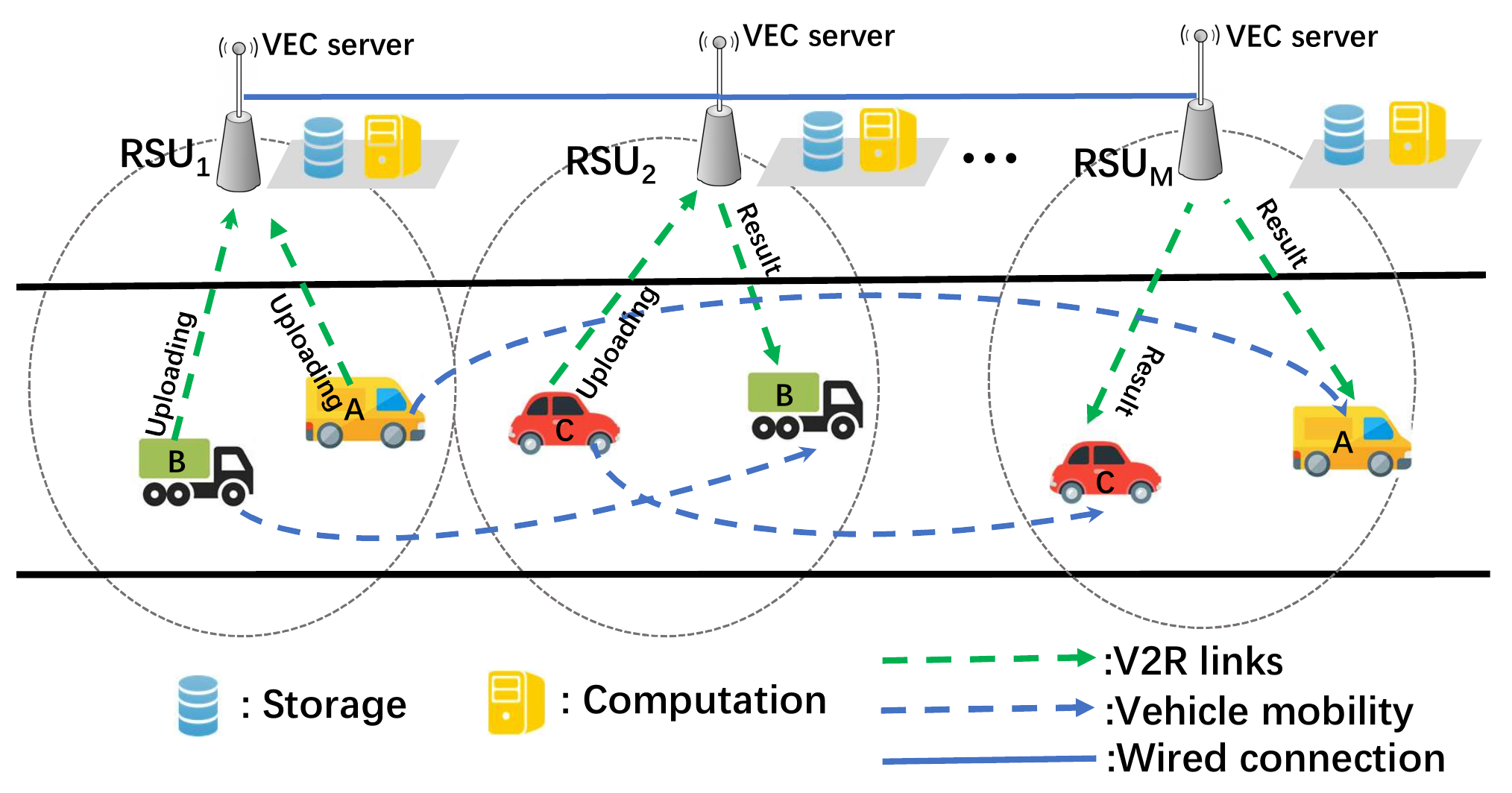}
  \caption{Example of mobility awareness}
  \label{f6}
\end{figure}

In VEC, edge servers are empowered by rich computation and storage resources in proximity of vehicular users. However, suffering the limited coverage range of edge servers (e.g., RSUs) and the high mobility of vehicles, there is a possibility that the task from one vehicle cannot be finished successfully within the original edge server. That is to say, multiple edge servers may be passed through before the task is accomplished. In order to deal with this issue, mobility-based task migration among different edge servers represents one attractive solution, which has gained increasing attention \cite{zhang2017mobile} \cite{dai2018joint2}. When a vehicle is approaching a new server, its task can be migrated to this server, or the task can be computed in the original server and then the obtained result is  forwarded to the vehicle via the new server. As shown in Fig. \ref{f6},  vehicle A  has a task for implementation. Thus, it uploads the task to the associated edge server $RSU_{1}$. Before the result is calculated, vehicle A has entered into the coverage range of $RSU_{m}$. In this situation, $RSU_{m}$ can obtain the result from  $RSU_{1}$, and then deliver the result to vehicle A.  By considering the requirements of computation tasks and the mobility of vehicles, a predictive-model transmission strategy is introduced in \cite{zhang2017mobile} for task offloading, improving the transmission efficiency and satisfying the required delay. Besides, an optimal  offloading scheme based on prediction is designed for serving diverse types of computation tasks.

If all the vehicles offload their tasks to the same edge server,  it is likely to be overloaded, which can degrade the network performance. The authors of \cite{dai2018joint2} investigate resource allocation by jointly considering load balancing and offloading in a multi-user multi-server vehicular environment. This proposed solution not only allows vehicles to select their preferable edge servers based on their requirements but also accounts for the mobility of vehicles.


\subsection{Incentive Strategy}

Because of the self-interested characteristic, existing entities in the network are unwilling to contribute their available resources freely to others. Thus, a  satisfied incentive scheme is the necessity of encouraging them to participate in resource sharing, i.e., they should be rewarded once sharing the resources. As described above, the work of \cite{li2018parked} exploits the resources of PVs by giving them necessary compensation.

 Contact theory has been widely applied in resource management owing to its benefit in making two rational entities of a trading to reach consensus. Thus, a contact theoretical approach is employed in \cite{zhang2017contract} to provide the optimal offloading strategy for an edge server by maximizing the revenue of the edge server while improving the utilities of vehicles. An efficient resource management algorithm is provided to optimize the resource utilization of edge servers by considering the distinction of task priorities as well as additional resources.
Stacklberg game is used in \cite{zhang2017optimal} to design an optimal multilevel offloading scheme to make the utilities of vehicles and edge servers maximum. 
A backup server in vicinity is utilized to supplement insufficient resources of edge servers, due to the fact that if edge servers are located in dense roads, their constrained capacities may negative impact the QoS of vehicular users.

\subsection{Energy Efficiency}

Constrained by the limited battery energy of vehicular users ( e.g., electrical vehicles, smart phones and wearable devices), the battery endurance time will be sharply reduced when they process tasks locally. Under this condition, VEC allows vehicular users to enlarge the battery life by offloading their energy-hungry tasks on the network edge with sufficient energy supply.
The work in \cite{zhou2018energy} devises  an energy-efficient VEC architecture to enable service provisioning for energy-constrained users. Based on two built  models with respect to energy consumption and delay, the problem of minimizing energy consumption is formulated as a joint workload offloading and power control problem. Unmanned Aerial Vehicles (UAVs) assisted by MEC are introduced in \cite{zhang2018energy} to enhance computation capacity. Because of the advantages of low cost and flexible deployment, UAVs can deal with the limited resources of vehicles as well as the considerable cost and huge burden of RSUs. The interesting idea of \cite{ku2018quality} is the introduction of Solar-powered RSUs (SRSU) to  VEC for providing vehicles more energy-efficient communication. The investigation aims to solve the gap between solar power generation and SRSU energy consumption for improving the QoS.

\subsection{AI-enabled Offloading}

There have been several work to apply AI into  task offloading in VEC. For example, machine learning is widely used \cite{zhu2018deep}.
In \cite{he2018integrated}, a resource allocation strategy by joint taking networking, caching and computing into account is formulated as an optimization problem. However, the joint consideration of these three factors increase the complexity of the problem. Thus, a deep reinforcement learning method is introduced to solve the optimization problem.
In \cite{tan2018mobility}, the joint communication, computation cache problem is studied. The resource scheduling is designed by taking into account the mobility of vehicles and the hard delay requirement. The deep learning algorithm is utilized to solve the resource allocation problem. In \cite{dai2018AI}, by utilizing deep reinforcement learning, the authors deal with joint edge computing and caching resource allocation problem. The authors of \cite{zhang2019deep} design a deep Q-learning approach for making optimal offloading decisions by taking into account the selection of target server and the determination of data transmission strategy simultaneously.

%

\section{Caching in Vehicular Edge Computing}\label{s5}

\begin{figure}[H]
  \centering
  \includegraphics[height=5cm, width=8cm]{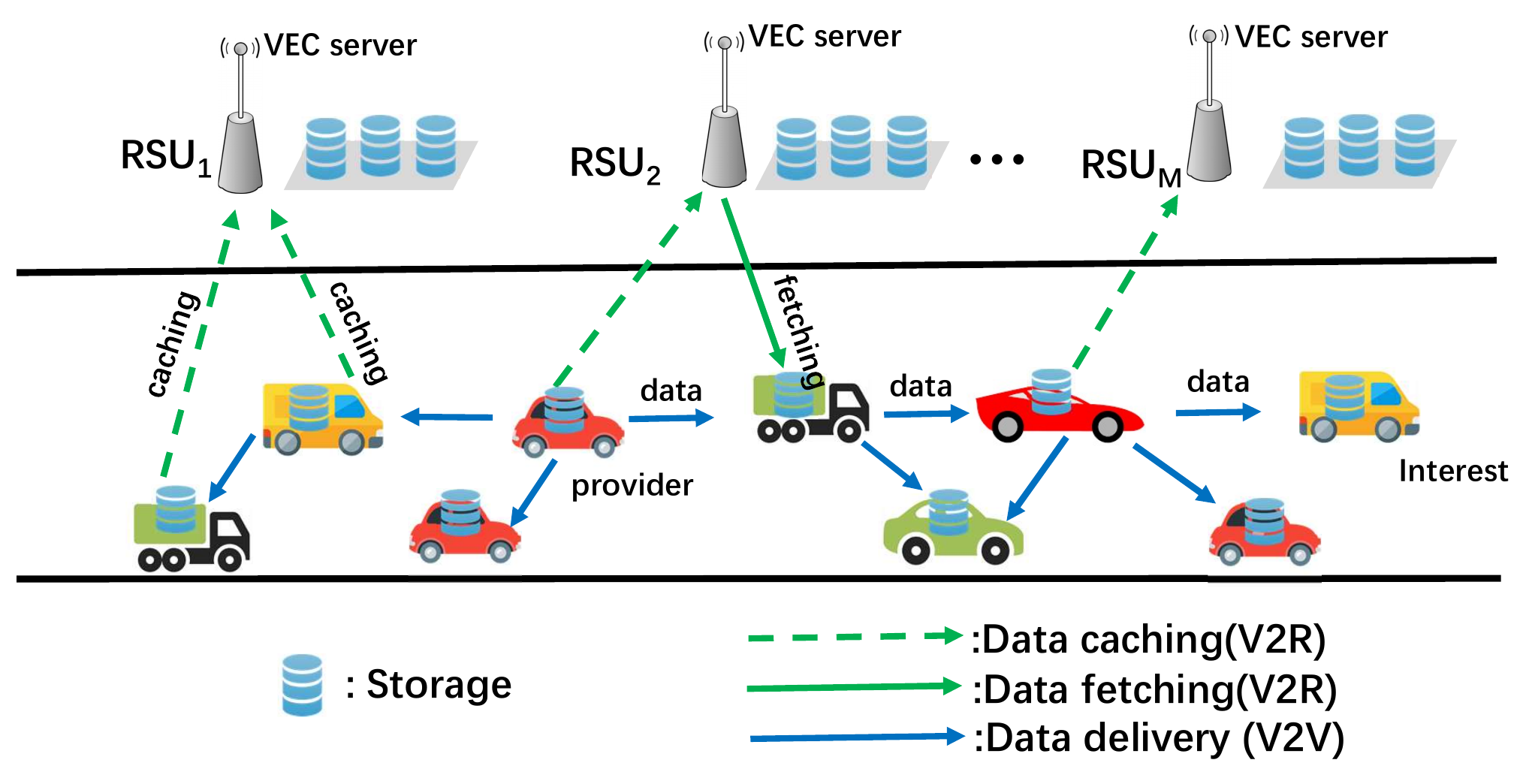}
  \caption{Example of caching }
  \label{f71}
\end{figure}

With the proliferation of smart vehicles, vehicular networks are undergoing substantial growth in data traffic due to the the boom in various data-consuming applications. This brings great challenges on backhual networks and latency. Edge caching is regarded as an attractive scheme to deal with these challenges \cite{lien2018energy}. Being supplement to the cloud, the edge network can be enriched with rich storage resources. As shown in Fig. \ref{f71}, edge servers and vehicles with ample storage resources can be utilized to cache popular contents. By caching contents in proximity of users, the contents can be accessed locally without resorting to the remote cloud, enabling the reduction of redundancy data traffic, the alleviation of scarce network bandwidth and the improvement of user experience.

Generally, there are three primary questions that need to be answered for content caching:

\begin{table*}

\caption{Summary of literatures on caching}
\begin{center}
\begin{tabular}{|p{3cm}|p{1cm}|p{3cm}|p{8cm}|}
\hline
\textbf{Theme} & \textbf{Reference} & \textbf{Metric}& \textbf{Key Points} \\ \hline
 \multirow{9}*{Caching at RSUs}&\cite{hu2017roadside}& Dissemination efficiency &Introduce a multi-object auction-based solution to deal with the competition among multiple content providers\\ \cline{2-4}
 &\cite{ding2015roadside}& Delay&Present three algorithms to allocate contents on RSUs\\ \cline{2-4}
 &\cite{su2018edge}&Hit ratio& Propose a dynamic content caching scheme based on cross-entropy\\ \cline{2-4}
 &\cite{mahmood2016mobility}&Probability of downloading contents& Design a probabilistic caching scheme with the consideration of vehicles' trajectories and content service time by edge nodes\\ \cline{2-4}
 &\cite{mahmood2018rich}&Probability of downloading contents&The probabilistic caching scheme accounts for the mobility of vehicles\\ \cline{2-4}
 &\cite{zhang2017cost}&Network deployment cost and QoS& Propose a cost-effective network planning assisted by cache-enabled green RSUs\\ \hline
 \multirow{6}*{Caching at vehicles}&\cite{caching}&Delay& The storage resources of PVs in multiple parking lots are exploited to cache contents from multiple content providers\\\cline{2-4}
&\cite{kumar2014peer}&Traffic load& Design a P2P cooperative caching scheme \\ \cline{2-4}
&\cite{fang2017cooperative} &Network cost& Propose a cooperative caching scheme by considering the global status of clusters as well as the constraints of vehicles with respect to caching capacity \\\cline{2-4}
&\cite{quan2016intelligent}&Efficiency of content retrieval& Propose a popularity-aware content caching and retrieving mechinasim\\ \cline{2-4}
&\cite{hu2018vehicle}&Retrieval outage and delay& Propose an in-vehicle caching scheme which has the advantage of flexibly accommodating multi-file popularity according to the designed dynamic distributed storage relay scheme\\ \cline{2-4}
&\cite{deng2016distributed}&Delay and caching utilization&Caching decisions of vehicles depend on the demands of users, the importance of vehicles as well as the characteristics of relative movement\\ \cline{2-4}
&\cite{yao2018cooperative} &Retrieval outage, delay and caching utilization& A cooperative caching scheme is designed based on the mobility prediction of nodes\\ \hline
\multirow{2}*{Multi-layer caching}&\cite{ma2017low}&Delay and QoE& The caching policy is designed by considering both of vehicular layer and RSU layer\\ \cline{2-4}
&\cite{zhang2018cooperative}&Delay&Vehicles are exploited as caching nodes for collaborative content caching with BSs\\ \hline
\multirow{4}*{AI-enabled caching}&\cite{ndikumana2018deep}&delay&MLP is utilized to predict the probability of contents being requested, and CNN is employed to predict the ages and genders of passengers\\ \cline{2-4}
&\cite{hou2018q}&delay& Present a Q-learning based caching scheme by integrating a deep learning network\\ \hline
\end{tabular}
\end{center}
\end{table*}


\begin{enumerate}
  \item Caching place:In vehicular edge network, there are two main edge devices, say, edge servers (e.g., RSUs) and vehicular users (e.g., vehicles), which are common places to cache contents;
  \item Caching contents:Because of the limited storage capacity of edge devices and the extreme amount of contents available over the entire network, it is not practical to cache all the contents for edge devices. In this situation, it is better to selectively cache the contents. By comparison, popularity-based cache placement represents a mainstream solution, where content popularity is an indicator of contents being requested by users;
  \item Caching policy:Caching policy depends on the motivation behind content caching, e.g., offloading traffic, reducing delay, improving QoE and minimizing energy consumption.
\end{enumerate}

In the following, the existing work related to edge caching is described by classification in detail. A summary of literatures on caching is illustrated in Table III.

%

\subsection{Caching at RSUs}

The explosion of emerging applications in vehicular networks, e.g., in-vehicle infotainment, poses tremendous challenges on the network capacity. Being important connected devices,  RSUs are seen as the significant infrastructures to be deployed for enhancing the network capacity \cite{cheng2011infotainment}, \cite{zhang2017self}. Though moving popular contents in close proximity to vehicular users, RSUs are capable of offering benefits in reducing delay and alleviating network burden.

Different from BSs, the storage capacity and communication range are limited, which will impact the caching decision. There is a massive amount of contents in vehicular networks, e.g., videos and audio files. It is crucial to make full use of the limited resource of RSUs and effectively decide the cached contents for improving the QoE of users.  In \cite{hu2017roadside}, multiple content providers intend to cache their contents themselves to facilitate data transmission. This leads to the competition from content providers for sharing limited resources of edge servers, thereby arising the issue of optimizing storage utilization of RSUs. A multi-object auction-based algorithm is used to solve the competition.

In vehicular networks, different content requestors may have different files. Thus, there are conflicts among them. How to place files in RSUs is extremely of importance. The study in\cite{ding2015roadside} focuses on distribution of popular contents among RSUs by minimizing the average time needed for content requestors obtaining their requested contents. The relationships between the average content downloading time with several parameters, including the number of RSUs, caching space and vehicle speed, are analyzed. In \cite{su2018edge}, A dynamic content caching scheme is designed to optimize caching through the requests of vehicles and the cooperation among RSUs. The features of content requests depend on content access pattern, vehicle mobility as well as density, by which a vehicle can decide whether and where to fetch the requested content. On the other hand, it is highlighted in \cite{idir2015optimal} that it will lead to an inefficient utilization of storage resources when all the RSUs are used to cache contents. Thus, a subset of RSUs are selected in \cite{idir2015optimal} to store data which are available for vehicles. An interesting point is that coding techniques allow the entire multimedia data being split into a series of small coded pieces. Vehicles can recover the entire data via any set of linearly independent codes pieces. The method can help in content transmission in dynamic vehicular networks.

High movement of vehicles in VEC gives rise to the varying network topology with time. The caused unstable connectivity makes it difficult to maintain efficient and continuous data transmission among nodes, negatively impacting the user experience \cite{chen2012available}. For example, it is likely  that vehicles cannot fetch the entire requested content in the coverage range of a RSU, and there is a possibility of passing though multiple RSUs for vehicles before obtaining the requested content. Thus, the mobility of vehicles is a prime concern when designing the caching scheme. A  probabilistic caching strategy is designed in \cite{mahmood2016mobility} with consideration of the prediction of vehicular trajectories and the content serve time at edge nodes, by which a mobility-aware caching scheme is proposed, aimed to optimally caching contents at edge nodes. The storage of content chunks on edge nodes in \cite{mahmood2018rich} is dependent on the historical statistics of achievable data rates and the standing time of vehicles within the coverage area of edge nodes.

The deployment of traditional RSUs relies on the support of power lines and wired backhuals. This is extremely challenging and costly in  areas where power systems are under-developed and infrastructures are sparse.  In order to tackle this issue, RSUs installed with solar panels or wind turbines are alternative. Such Green RSUs have the ability of harvesting renewable energy as power source. This type of cache-enabled green RSUs is utilized in \cite{zhang2017cost}, which can facilitate flexible deployment, energy-saving operation as well as low delay services. Assisted by these RSUs, a cost-effective network planning is designed to reduce the network cost while achieving desired data rate.

\subsection{Caching at Vehicles}

 Enabled by rich storage resources, smart vehicles can be employed as special infrastructures to make the full utilization of these resources owned by them. Cooperative caching allows vehicles to achieve desirable hit ratio and reduced delay.  In urban environments, storage resources can be greatly improved when vehicles share their unused storage space in a collaborate way. A typical example is PVs. Benefiting from the large number and long dwell time, PVs are ideal choice for storage provisioning.
In\cite{caching}, the storage of PVs in multiple parking lots are used to cache popular contents provided by content providers in a cooperative manner. 

Although the existing fixed infrastructures, e.g., RSUs, can provide connectivity to vehicles for content access, suffering from the high mobility and sparse distribution of vehicles, it is not an ideal option to only cache contents in the existing infrastructures deployed in specified places. In this situation, a cooperative caching scheme is useful for data dissemination \cite{kumar2014peer}. In \cite{kumar2014peer},  traffic information is shared in a Peer-to-Peer (P2P) manner among vehicles. The previous cached data is updated by newly arrived data using a probabilistic manner.

The caching at vehicles also faces several challenges with respect to limited storage capacity, popularity calculation and vehicle movement. The cooperative caching algorithm designed in \cite{fang2017cooperative} considers the global caching status of  clusters and the limited caching capability of vehicles. In \cite{quan2016intelligent}, the calculation of popularity is based on  the reference frequency of content and chunk. Based on the popularity of contents, vehicles can switch the retrieval models between V2V and V2R. The designed caching scheme will be triggered once the interest increasing ratio of a requested content is too big, for the purpose of alleviating sudden huge requests. The work of \cite{hu2018vehicle} provides a dynamic distributed data storage scheme by  explicitly accounting for the mobility of vehicles, the key core of which is to maintain the survival of storage data in a designated region of interest by means of forwarding the data from the leaving vehicles to the arriving vehicles. Based on the novel data storage scheme, the designed in-vehicle caching strategy can support multi-file popularity flexibly. In \cite{deng2016distributed}, a distributed probabilistic caching strategy is designed, where each node makes caching decisions by considering the demands of vehicles, the importance of vehicles and the relative movement of vehicles. 
The authors in \cite{yao2018cooperative} focus on the selection of caching nodes. In this work, based on the historical trajectories of vehicles, the probability of vehicles to visit different hot spot regions can be predicted. The vehicle which stays longer in a hot spot region is preferable to be caching node.

\subsection{Multi-layer Caching}

In VEC, in order to improve resource utilization, any devices with storage space are potential caching nodes. In other words, contents are allowed to be cached in at different layers, including BSs, RSUs and vehicles if they have storage resources. Caching contents at different layers is beneficial for increasing the availability of contents being successfully accessed by users on move \cite{kumar2015qos}. Thus, an advanced caching policy should take full advantage of different layers to achieve optimal performance. In \cite{ma2017low}, the caching placement is investigated to take into account both of vehicular layer and RSU layer, aimed to minimize the delay while satisfying the QoE of users. The work \cite{zhang2018cooperative} investigates a mobility-aware caching framework for vehicular uses, where vehicles are used to provide content sharing with Base Stations (BSs). Frequent disconnections caused by high mobility of vehicles make it difficult to provide and maintain the QoS of video streaming applications. Thus, the proposed caching scheme  in \cite{kumar2015qos} considers both of the mobility of nodes and the link disconnection. In this proposed scheme, two metrics of Load Utilization Ratio (LUR) and Query to Connectivity Ratio (QCR) are designed to maintain the QoS.

\subsection{AI-enabled Caching}

In order to decide what to cache, the content popularity is taken into account by a lot of work. However, the mobility of vehicles will lead to the varying membership of vehicular users associated with individual edge nodes. Besides, the user interests in contents are changing with time in different contexts, e.g., location, network topology, as well as personal status. This makes content popularity unknown before performing caching operation \cite{zhu2018deep}. The authors of \cite{ndikumana2018deep} try to solve this issue by employing deep learning. In \cite{ndikumana2018deep},  the probability prediction of requested content is made by using a Multi-Layer Perception (MLP) algorithm. Meanwhile,  Conventional Neural Network (CNN) is utilized to predict the interest of passenger. By comparing CNN output with MLP output, the contents needed to be downloaded from edge servers for caching can be identified.

In order to improve caching services, proactive caching is widely used. However, due to the limited capacity of RSUs, the design of an optimal caching strategy to balance the QoS of users and the caching cost of RSUs is challenging. A Q-learning based caching scheme by combining a deep learning network is devised in \cite{hou2018q}  to solve this issue. In this scheme, the mobility of each vehicle is predicted using a long short-term memory network. Based on the predictions, an optimal proactive caching scheme is developed.

\section{Data Management in Vehicular Edge Computing}\label{s6}

The significant increase of various types of application services leads to the exponential growth of data. From one hand, the installed devices in vehicles, such as GPS, camera and larder, can generate a multitude of dada; from the other hand, data (e.g., video streaming) can be shared among vehicles via wireless communication manner. This brings huge challenges on response time and bandwidth of backhaual networks for data processing. In this situation, it is a major concern to  efficiently manage data, including data collecting, analyzing, processing and dissemination. To the end,  there have been plenty of work done for data management. In the following,  some of them are introduced by classification. A summary of literatures on data management is illustrated in Table IV.

\begin{table*}

\caption{Summary of literatures on data management}
\begin{center}
\begin{tabular}{|p{4cm}|p{1cm}|p{3cm}|p{8cm}|}
\hline
\textbf{Theme} & \textbf{Reference} & \textbf{Objective}& \textbf{Key Points} \\ \hline
 \multirow{4}*{Data Collecting and Processing}&\cite{lai2019efficient}& Request Answering&Propose a filter-based request answering framework\\ \cline{2-4}
&\cite{hagenauer2017vehicular}& Data collection & Vehicular virtual cloud is introduced as virtual edge server\\ \cline{2-4}
&\cite{lai2018fog}& Data monitoring and gathering & A fog-based two-level threshold scheme is designed for preventing the transmissions of unnecessary data \\ \cline{2-4}
&\cite{darwish2018fog}& Data analytics & Present a real-time big data analytics framework\\ \hline
\multirow{3}*{Data Dissemination}&\cite{hou2017design}&Message dissemination&Design a message dissemination scheme assisted by MEC\\ \cline{2-4}
&\cite{iqbal2018context}&Context-aware service& Propose a context-aware data-driven intelligent framework\\ \cline{2-4}
&\cite{kadhim2019energy}  &Multicast routing& Devise an energy-efficient multicast routing scheme with SDN and fog computing\\ \hline
\multirow{3}*{Content Delivery}&\cite{jiao2018proactive}& Content delivery & Study the tradeoff problem of communication, computing and cache for proactive caching \\ \cline{2-4}
&\cite{huang2016reliable}&realtime streaming&The streaming contents are allocated in advance from computing service providers \\ \cline{2-4}
& \cite{luo2018cooperative}& Content distribution & Investigate the prefetching and distribution of contents  \\ \hline
\multirow{2}*{Incentive Scheme}&\cite{hui2018content} &Content dissemination& Because of the selfishness and capacity of vehicles, a relay selection scheme is provided to select vehicles to relay the contents\\\cline{2-4}
&\cite{magaia2018repsys}&Caching and dissemination& Propose a robust and distributed incentive mechanism for content caching and dissemination in a collaborative method\\ \hline
\multirow{5}*{Performance Optimization}&\cite{gangadharan2018bandwidth}&Bandwidth optimization& A vehicle flow model is introduced into the optimization framework\\ \cline{2-4}
&\cite{yaqoob2018fog}&Congestion avoidance& Present a fog-based congestion avoidance strategy \\\cline{2-4}
&\cite{chen2017exploring} &Data scheduling& Design two dynamic scheduling algorithms to schedule data \\ \cline{2-4}
&\cite{zhou2018begin}&Data processing&Propose a cooperative fog computing mechanism for big data processing\\ \cline{2-4}
&\cite{zhang2017cooperative}&Energy-efficient&Integrate big data analytical into VEC\\ \hline
\end{tabular}
\end{center}
\end{table*}


\subsection{Data Collecting and Processing}

A large amount of data are generated every day. Each vehicle is a data producer as well as a consumer in vehicular networks. These data should be collected and processed well. If these data are be deeply exploited, more learned knowledge can be used to facilitate the data efficiency and improve the network performance.

A filter-based framework is provided in \cite{lai2019efficient} by integrating fog computing with vehicular sensing. This framework makes full use of pull and push schemes to collect the requested data in vehicular networks in an adaptive and efficient manner. The authors of \cite{hagenauer2017vehicular} introduce the other data collection approach. The concept of vehicular micro clouds as virtual edge servers is established to optimize data processing and aggregation before delivering data to a data center. 
The problem of unnecessary data transmission is solved in \cite{lai2018fog} by the introduced fog-based two-level threshold approach. From one hand, the presented scheme can adaptively adjust the threshold to unload a suitable volume of data for decision  making; from the other hand, it can prevent unnecessary data transmission. The realtime big data analytics in vehicular networks is investigated in \cite{darwish2018fog}, by combining  intelligent computing and real time big data analytics.

\subsection{Data Sharing}

\subsubsection{Data Dissemination}

In vehicular networks, delay-sensitive applications account for a large proportion. These applications have stringent delay requirement.  An efficient data transmission scheme is crucial to facilitate the reduction of delay and the improvement of reliability \cite{liu2018data}. For example, safety message needs to be forwarded quickly and precisely to enhance road safety and avoid potential danger.  With the aid of MEC, a publish/subscribe message dissemination scheme is developed in \cite{hou2017design}. This designed scheme is composed of three important roles. The first one is message broker which supports message services, the second one is message hander aimed to explore potential data value, and the last one is message hub integrating different parts of the system. In this scheme, MEC is used to provide diverse application services, especially those with critical latency requirement. The article in \cite{iqbal2018context} provides a data analytics framework driven by the faced challenges  when offering context-aware services in vehicular networks, which enables delay-critical applications by enhancing the fog layer of traditional vehicular network architecture. A multicast data dissemination protocol is proposed in \cite{kadhim2019energy} to reduce the energy consumption with the constraints of bandwidth and deadline. 
Classification algorithm is designed to make the classification of received multicast requests according to different application types, while scheduling algorithm is introduced to manage multicast requests on a basis of priorities. Moreover, a partitioning scheme is employed to reduce the overhead and time complexity of the proposed algorithm.

\subsubsection{Content Delivery}

The explosive growth of vehicles gives rise of high amount of content requests. The use of content delivery can alleviate the huge network burden as well as reduce the delay. If the requested contents are stored in some nodes, by depending on the content delivery among nodes, there is no need to access the remote cloud for content acquiring, shortening the delivery time. The work of \cite{jiao2018proactive} provides a theoretical architecture to enable content delivery  by balancing computation, caching and communication resources. A concrete distributed scheme for reliably delivering realtime contents is presented in \cite{huang2016reliable}
. The core of this scheme lies in that the streaming contents are allocated in advance from cloud and fog  to guarantee the QoS of realtime streaming. By predicting the possible locations, the amount of required contents  as well the needed tokens, mobile nodes can reserve contents for streaming, ensuring the reliability of streaming content. Besides, this scheme is capable of effectively adapting to the changing behaviors of mobile nodes as well as the condition regarding computation resources. Distinguished with \cite{huang2016reliable}, both of content prefetching and distribution are studied in \cite{luo2018cooperative}. A hierarchial architecture based on edge computing is introduced for efficiently facilitate the distribution of large-volume data. In order to deal with the challenges of dynamic topology change and unbalanced traffic,  a multi-place and multi-factor prefecting mechanism is proposed. The problem of content distribution is solved by using a game theory approach.

\subsection{Incentive Scheme}

Nodes with rich contents have no obligation to share their resources in vehicular networks. The selfishness of nodes is not considered in these work above. In order to encourage data sharing, the necessary incentive scheme needs to be designed well \cite{shi2018incentive}. This issue is investigated in \cite{hui2018content}, where  a content delivery solution based on edge computing is provided. In this scheme, contents need to be transmitted to an edge computing device
. The edge computing device makes the selection of available vehicles to deliver the contents 
with the joint consideration of  selfishness and capacity of vehicles. The selected vehicles are responsible for relaying the contents to the vehicles which have interest in the contents on the way to the destinations. The work in \cite{magaia2018repsys} provides an incentive approach for collaborative caching and dissemination in vehicular delay-tolerant networks (VDTNs). This approach has two distinct benefits: robust and distributed. It performs robust because of the resilience against false accusation and praise, and is distributed based on the fact that the decision made to interact with another node should reply on each node. 

\subsection{Performance Optimization}

The performance metrics, including delay, energy, bandwidth, reliability, etc.,  are main indicators of the network performance, which provide the guidance of optimizing the network performance. In \cite{gangadharan2018bandwidth}, with the objective of bandwidth optimization, a centralized optimization framework is introduced to deliver data and provide services to vehicles. The framework  takes into consideration the edge resource constraints and integrates the vehicle flow model simultaneously. A fog-based congestion avoidance scheme 
is proposed in \cite{yaqoob2018fog}, aimed to alleviating the congestion and reducing the delay. In this scheme, a fog server is employed to conduct services in vehicular networks. In order to alleviate the pressure of the network load brought by the increasing data disseminations, fog computing is utilized in \cite{chen2017exploring}. The wide deployment of edge computing infrastructures will lead to the issues of energy consumption and air pollution, which motivates the design of a programmable, scalable and flexible architecture in \cite{zhou2018begin} by the integration of big data analytics in VEC. The authors of \cite{zhang2017cooperative} present a regional cooperative fog computing to provide diverse services. A localized coordinator is utilized to support interoperability and cooperative operation among local fog servers. A hierarchial resource management model is developed  to optimize the network performance in the fog computing network. Two scheduling schemes on a basis of response time and queue length are presented to schedule data for adapting the changing network and enhancing the efficiency of data dissemination, respectively.

\section{Flexible Network Management in Vehicular Edge Computing}\label{s7}

It is mentioned in \cite{zhang2018software} that the key of next generation vehicular networks lies in the design of heterogeneous networks using advanced technologies, such as FC and SDN. 
SDN has the feature of enabling  flexible and dynamic network management by separating  control plane and data plane. When receiving the commands from the controller deployed in the control plane, the data plane conducts the corresponding operations. SDN benefits the reduction of network management costs and is able to get a global knowledge of networks via the collected information from the data plane. The integration of VEC with SDN can provide more efficient and reliable centralized control management. The recent research regarding software-defined vehicular networks is summarized in \cite{deng2017latency}, where many edge-up software-defined networking decisions are proposed with a special focus of delay control for providing various services in vehicular networks. However, because of the diversity of fog applications, the design of an efficient vehicular network architecture is still an open problem. The authors in \cite{nobre2019vehicular} make an attempt to study the design principles for fog-enabled vehicular SDN, mainly emphasizing the perspectives of systems, networking as well as services. 
Next, some efforts regarding the architecture design with the combination of SDN and VEC in the literature are introduced.

A vehicular architecture based on SDN and FC  is developed in \cite{zhang2018software} for satisfying the requirements of users in terms of low delay and high reliability. In this architecture, the mobility management is analyzed with a focus of handover management mechanism. Then, a hybrid access handover strategy for improving the handover performance and a resource allocation scheme for optimizing the delay are proposed, respectively.

A SDN-enabled heterogeneous vehicular network architecture assisted by MEC is proposed in \cite{liu2017scalable}. This architecture can ensure  required data ratios and good reliability in vehicular communication, and achieve desired scalability and responsiveness as well. The other network architecture by combining SDN with MEC is introduced in \cite{huang2017exploring} for 5G-enabled Software-Defined Vehicular Networks (5G-SDVNs). In this architecture, SDN  brings the benefit of improving network management and MEC is used to enhance the network control. This architecture can achieve significant performance improvements in terms of network management, resource utilization and network development.

 The designed vehicular network architecture in \cite{li2018delay} aims to alleviate the traffic congestion by jointly optimizing networking, storage and computation resources. The programmable control principle of SDN is introduced to the  architecture to improve the network optimization and resource management.

The presented architecture of 5G vehicular networks in \cite{soua2018multi} combines SDN, cloud computing and fog computing. In this architecture, vehicles serve as fog infrastructures, by which the network performance can be enhanced.

In the architecture introduced in \cite{choo2018optimal}, the controller is responsible for determining the task offloading strategy of vehicles and resource allocation strategy of edge cloud. It is crucial to select the edge cloud and  make the resource allocation to maximize the probability of a task being successfully finished with a given period. 

The work in \cite{kadhim2018maximizing} tries to optimize the utilization of fog computing resource, with the aim to fulfill the delay requirement of tasks by assigning each task with a best fog server with the assistance of SDN. 

\section{Security and Privacy in Vehicular Edge Computing}\label{s8}

With the advent of VEC, security and privacy are main concerns, which will impact the wide deployment and fast development of VEC. In VEC, the high mobility of vehicles makes it challenging to solve trustworthy problem among nodes. Because of the fact that a large number of different types of devices coexist in vehicular networks, the generated heterogeneity makes  traditional trust and authentication schemes inefficient.
Despite of the merit of moving the cloud to close proximity of users, edge servers are the potential target due to the public deployment without any physical isolation and also have the possibility of turning into a malicious because of the curious for user information.
However, the research on security and privacy in VEC is still at initial stage with a lot of problems unsolved. On the other hand, this  provides many attractive research opportunities for researches from academia and industry. Actually, there have been significant efforts devoted to this field with some introduced as follows.

\subsection{Security}

\subsubsection{Trust and Authentication}
Trust is of vital importance in security, the basic idea behind which is to know the identity of an entity with whom there is an interaction. Authentication management represents one possible approach to provide the guarantee of trust \cite{roman2018mobile}. The explosive growth of vehicles increases the number of misbehavior vehicles. These misbehavior vehicles can disrupt the VEC and reduce the network efficiency. Reputation management is seen as one key metric to access trustworthiness of vehicles \cite{ma2018privacy},\cite{ma2018robust}. By rewarding cooperative vehicles but punishing mishehavior vehicles, reputation management is of significant benefit in guiding user behaviors, preventing possible attacks as well as improving overall network performance.  In \cite{huang12017distributed}, reputation management is investigated to provide secure protection and promote network efficiency. In the proposed system, edge servers are responsible for carrying out local reputation management tasks. 
Reputation values of vehicles can also provide the guidance of optimizing resource allocation. Different from \cite{huang12017distributed}, to deal with untrustworthy nodes, negative messages are used to indicate negative attributes of such nodes in \cite{huang2017meet}. A scheme is provided to efficiently distribute these negative messages in vehicular networks, where RUSs act as the functionality of fog. Besides malicious attackers and faulty nodes which bring huge security thereat, the widely existing obstacles, e.g., buildings, trees and big trucks, will obstruct the vision and communication Line of Sight (LOS) of drivers, creating negative impact on the integrity and  reliability of data as well as availability of localization service. This issue motivates the proposal of a fuzzy trust model on a basis of experience and plausibility in \cite{soleymani2017secure}, where many security checks are carried out to guarantee the correctness of received information from authorized vehicles and fog nodes serve to assess the accuracy of event location.

\subsubsection{Confidentiality and Integrity}

In VEC, being the supplement of a remote cloud, RSUs with powerful computation capacity can undertake the tasks from vehicular users to improve the user experience. However, because they are regularly deployed at public places, there is the risk of being eavesdropped. After locating a RUS by using traffic statistics, the adversary can launch an attack on the RSU and interrupt on-gonging services. In order to avoid this issue,  the work in \cite{huang2018secure} introduces one proactive defend solution depending on generating dummy traffic delivery. The generated dummy packets by vehicles can be used to provide protection for RSUs by misleading the traffic statistics.

The data offloaded to RSUs may be sensitive and even private information, such as location, purchase history and healthcare record. Thus, in order to avoid information leakage, some necessary measures should be taken. The study in \cite{wu2018secrecy} targets the confronted eavesdropping attack in case that vehicles offload their tasks to the network edge via Radio Frequency (RF) channels. The resource management is investigated for secrecy provisioning based on a tool of physical layer security. The authors of \cite{yao2018reliable} focus on the security of data delivery. Vehicular Fog Service (VFS) is provided by exploiting the computation and storage resources of PVs. A mechanism composed of a vehicular fog construction approach and a VFS access approach is used to guarantee the reliability and security of VFS. 
In \cite{chen2018privacy}, 
data collected by vehicles can be sent to RSUs in a privacy-preserved manner. The suggested certificateless aggregate signcryption mechanism (CL-A-SC) can support certificateless cryptograph and signcryption at the same time. Being the extension of CL-A-SC, the proposed anonymous aggregation protocol helps in obtaining good security properties.

 Although cryptography-based schemes have the benefit in dealing with security and privacy, resource-constrained vehicles cannot carry out them due to the intensive computation requirement. In \cite{xue2018fog},  the complicated encryption and decryption tasks are migrated to fog servers and cloud server with the preservation of confidentiality and privacy. 
Besides, the designed interactive protocol is capable of providing verifiable recording of fog servers' reports while preserving users' anonymity and unlinkability.

\subsection{Privacy}


Real-time  navigation system is beneficial for finding an optimal route by the collected realtime traffic information. Before implementing this system, security and privacy should be addressed well. To the end, a secure and privacy-preserving navigation scheme is proposed in \cite{wang2017secure}.

Road condition is regarded as one key metric of reflecting the quality of roads. Thus, it is crucial to monitor road conditions for enhancing road safety. By using the sensing capacity of mobile devices without the deployment of additional sensor nodes, the sensory data can be collected to improve the transportation system, forming a new sending paradigm named as mobile crowd sensing. However, safety and privacy issues emerge during the process of data processing.  In \cite{wei2018privacy}, a fog-based privacy-preserving mechanism is designed to ensure the security of vehicular crowd sensing networks. A certificateless aggregate signcryption algorithm is proposed in \cite{basudan2017privacy}. Based on the algorithm,  a data transmission scheme is presented for monitoring road conditions to guarantee the requirements in terms of information confidential, integrity, authenticity, privacy and anonymity. 




Location privacy has been becoming more and more important in vehicular networks due to a large amount of data sharing, especially with the emergence of location sharing applications, e.g., Google Maps. A vehicle can be easily tracked by its adversary based on its communication and movement behaviors. Vehicles need to send their driving information to neighbors for safe driving and expose their location information because of the utilization of location-based services. An adversary is able of locating the vehicle' position once obtaining these information. 
The work in \cite{li2018pros} makes an attempt to address the user privacy issue stemming from route sharing by introducing a privacy-preserved route-sharing scheme. In order to deal with location privacy, the real identity of vehicles can be hidden using pseudonym technology. The authors of \cite{kang2018privacy} provide a privacy-preserved pseudonym to protect location privacy. Pseudonym fogs consisting of a group of fog infrastructures are responsible for pseudonym management of vehicles passing by. Based on the context information provided by local authorities deployed in pseudonym fogs, vehicles can conduct context-aware pseudonym changing. In order to avoid the problem of a completely anonymized vehicle turning into a malicious, conditional privacy is needed to identify and revoke misbehaving vehicles. In \cite{huang2011pacp}, conditional privacy is provided to allow vehicles to be anonymous until they are identified and revoked when they turn into malicious. The proposed privacy preservation protocol permits vehicles to use pseudonyms for privacy protection. For the sake of anonymous communication, vehicles get the pseudonyms by interacting with RSUs. During this process, only vehicles know their pseudonyms. In addition to pseudonym technology, a differentially location privacy-preserving scheme is designed in \cite{zhou2018achieving} with the aim to guarantee location privacy. When a vehicle sends the request of location-based service, it needs to send its location to the edge server, then the edge server executes the privacy-preserving scheme to query the Point of Interest (POI) and filter the useless results. This designed scheme can make vehicles to get useful information based on the submitted location with no need to reveal their privacy location. In \cite{arif2018secure}, a query-based location and communication scheme is provided to using a fog server with fog anonymizer. The provided protocol can preserve location privacy well.

\subsection{Blockchain-enabled Protection Scheme}

Benefiting from the characteristics of blockchain including decentralization, anonymity and trust, many researchers have devoted their efforts in integrating blockchain into VEC for security and privacy protection \cite{kang2018blockchain}, \cite{dai2019blockchain}.

In order to ensure secure data storage and sharing in VEC, two technologies of consortium blockchain and smart contact are fully exploited in \cite{kang2018blockchain}, which can efficiently avoid unauthorized data sharing. Then, a data sharing strategy based on reputation is provided to achieve high-quality data sharing, where a three-weight subjective model is used to make precise reputation management among vehicles. The proposed system of enhancing security protection in \cite{huang2017distributed} is also related to reputation management. In this system, local reputation management tasks are conducted by the servers. To enable the improvement of network performance, this system includes the following outstanding characteristics: distributed reputation management, trusted reputation manifestation, accurate reputation update as well as available reputation usage. 
The optimization of resource allocation is made by service providers in computation offloading to enhance reputation usage with the consideration of reputation of vehicles.
The work in \cite{kang2018towards} develops a secure P2P data sharing strategy, where the improved delegated proof-of-stake consensus scheme is employed as a hard security solution.  This delegated proof-of-stake consensus scheme is enhanced by a two-stage soft security approach: miner selection and block verification. The former is used to select miners through reputation-based voting, where the reputation of each miner candidate is calculated by a multi-weight subjective logic scheme; while the latter aims to incentive the standby miners with high reputation to take part in block verification.

Assisted by this presented blockchain-based decentralized trust management system in \cite{yang2018blockchain}, vehicles can assess the received messages from neighbors. With the result, the vehicle is able to create a rating for each source vehicle generating the message. Based on the ratings sent by vehicles, RSUs can calculate the trust value offsets of involved vehicles. Using blockchain technique, all the RUSs work with cooperation to build a reliable and consistence database.

In this provided privacy-preserving carpooling scheme in \cite{li2018efficient}, users are authenticated in a conditional anonymous manner. Private proximity test is used to realize one-to-many matching, which is extended to establish a unique key between a passenger and a driver. With a range query technique, get-off location matching can be achieved. A private blockchain is built in the carplloing system for the storage of carplloing records.

\section{Open Research Issues and Future Work}\label{s10}

The VEC has significant potential in improving the current transportation systems by integrating MEC into vehicular networks. However, the research on VEC is still in early stage. Many issues remain to be solved well. In this section, we identify some open research issues which can be extended to future research directions.

%

%

\subsection{QoS}

There are a variety of applications in vehicular networks, which are mainly classified into safety applications and non-safety applications.  The QoS demands of different types of applications are expected to be different. Safety applications (e.g., collision avoidance and traffic control)  have strict delay requirement, which should be addressed as soon as possible, while several non-safety applications (e.g., multimedia downloading) can tolerant some delay.  Thus, providing a flexible scheduling scheme to guarantee the QoS of different applications based on their priorities is a matter of concern in VEC.

\subsection{Scalability}

A massive amount of applications are increasing rapidly. This gives arise of large requirements for low delay, high reliability, rich computational capacity and storage space.  It is crucial to schedule resources optimally and conduct connectivity management efficiently for executing heterogenous task types. In addition, different from traditional cloud, vehicular users in VEC may be uneven distributed in vehicular networks. The densities of vehicles are varying with time in different areas. The designed system should adapt to the varying network conditions.

%

\subsection{Economic Profit}

The core of implementing the VEC lies in resource sharing. The resource owners are willing to share their resources  if they can be rewarded properly. In this situation, the pricing mechanism becomes extremely vital. This involves how to quantify the values of different resources to balance the profits of both resource consumers and resource providers. For resource providers, it should be considered to how to divide the retained profits among the related entities including the mobile service provider, edge service provider and cloud service provider.

\subsection{Security and Privacy}

Pushing the tasks with incentive computation and strict delay to the edge of networks is main application in VEC. The tasks offloaded to  edge servers usually includes sensitive and private data. In order to avoid the information leakage, data confidentiality should be ensured. In addition, with the aim to prevent the modification of data which are being forwarded or have been stored, it is extremely important to guarantee the integrity requirement. Meanwhile, a verifiable computing scheme for vehicular users is needed to demonstrate the correctness of obtained computation results from edge servers.

%

\section{Conclusion}\label{s11}

By the seamless merging of vehicular networks and MEC, VEC extends the function of cloud to the edge of networks, bringing rich resources in close proximity to vehicles. VEC is seen as an attractive network paradigm to fulfill the increasing requirements in terms of computation and storage resources. Different from traditional MEC, VEC is faced with several new challenges because of the intrinsic features of vehicular networks, including fast mobility of vehicles and harsh channel environments. 
Although many efforts have been done in VEC, there still lacks a survey on the research of VEC, which is the motivation of this paper.

In this paper, we present a comprehensive survey on the existing work in VEC. To this end, we first elaborate an overview of VEC consisting of the introduction, architecture, key enablers, advantages, challenges and application scenarios. Then, different VEC research topics are introduced. After that, we give a careful literature review on recent research by classification including task offloading, caching, data sharing, flexible network management as well as security and privacy. Finally, we identify several open issues and discuss future research directions.
The research on VEC is still on infancy with a lot of questions to be solved. It is expected that more efforts can be devoted into the new research field of VEC.

\bibliographystyle{IEEEtran}
\bibliography{IEEEabrv,VEC}

\end{document}